# Título: Calidad en repositorios digitales en Argentina, estudio comparativo y cualitativo

Autores: José Federico Medrano

Afiliación: Universidad Nacional de Jujuy / Universidad de Salamanca

Contacto: jfedericomedrano@gmail.com , Italo Palanca Nº 10, Facultad de Ingeniería, UNJu, San Salvador de Jujuy, C.P.: 4600

Resumen biográfico: Soy Ingeniero Informático por la Universidad Nacional de Jujuy (UNJu) y poseo una Maestría en Sistemas Inteligentes por la Universidad de Salamanca (España). Actualmente soy docente adjunto en la Facultad de Ingeniería-UNJu en las cátedras de Análisis y Diseño de Sistemas, Programación Aplicada y Programación Concurrente, además curso a distancia el Doctorado en Informática y Automática del Departamento de Informática y Automática dependiente de la Universidad de Salamanca. Trabajo como Administrador de Bases de Datos para el ministerio de Hacienda de la Provincia de Jujuy, en la Dirección Provincial de Rentas. Desarrollo software desde hace años y soy un apasionado de la algoritmia y las nuevas tecnologías. Actualmente desarrollo mis líneas de investigación en las áreas de Recuperación y Visualización de Información, sobre todo llevando estudios y análisis bibliométricos tomando como fuentes de datos motores académicos de libre acceso.

Resumen:

Son numerosas las instituciones y entidades que necesitan no solo preservar el material y las publicaciones que producen, sino también, estas tienen como tarea (sería deseable que sea una obligación) publicar, divulgar y poner a disposición del público los resultados de la investigación y cualquier otro material científico-académico. Para este propósito existen los repositorios de libre acceso, que a través de iniciativas como la *Open Archives Initiative* (OAI) y de la aparición de instrumentos como el protocolo *Open Archives Initiative Protocol for Metada Harvesting* (OAI-PMH), facilitan esta tarea en gran medida. El objetivo principal de este trabajo es realizar un estudio comparativo y cualitativo de los datos, específicamente los metadatos, contenidos en el conjunto total de repositorios argentinos que se encuentran listados en el portal ROAR[1], centrándose en la perspectiva funcional de la calidad de dichos metadatos, otro objetivo es ofrecer un panorama general del estado de dichos repositorios en un intento de detectar las faltas y errores comunes que incurren las instituciones al almacenar los metadatos de los recursos contenidos en estos repositorios y así poder sugerir medidas tendientes a mejorar los procesos de carga y posterior recuperación. Se encontró que los ocho campos Dublin Core más utilizados son: *identifier, type, title, date, subject, creator, language* y *description*. No todos los repositories cumplimentan todos los campos, además la falta de normalización o el uso desmedido de campos como language, type, format y subject es un tanto llamativa y en algunos casos alarmante.

---

[1] http://roar.eprints.org/

Abstract:

Numerous institutions and organizations need not only to preserve the material and publications they produce, but also have as their task (although it would be desirable it was an obligation) to publish, disseminate and make publicly available all the results of the research and any other scientific/academic material. The Open Archives Initiative (OAI) and the introduction of Open Archives Initiative Protocol for Metadata Harvesting (OAI-PMH), make this task much easier. The main objective of this work is to make a comparative and qualitative study of the data -metadata specifically- contained in the whole set of Argentine repositories listed in the ROAR[2] portal, focusing on the functional perspective of the quality of this metadata. Another objective is to offer an overview of the status of these repositories, in an attempt to detect common failures and errors institutions incur when storing the metadata of the resources contained in these repositories, and thus be able to suggest measures to be able to improve the load and further retrieval processes. It was found that the eight most used Dublin Core fields are: identifier, type, title, date, subject, creator, language and description. Not all repositories fill all the fields, and the lack of normalization, or the excessive use of fields like language, type, format and subject is somewhat striking, and in some cases even alarming.

Introducción

La sociedad de la información se ha convertido en los últimos años en la sociedad de la información abierta. Como lo indica (Méndez, 2015)... "Casi todos los conceptos de moda en nuestro entorno profesional incluyen, de una u otra manera la palabra abierto: *open access*, *open content*, *open data*, *open research*, *open education*, *open innovation*, *open knowledge*, etc.". Si bien no existe un consenso claro sobre lo que se considera conocimiento abierto, siguiendo la definición de *opendefinition.org*: "El conocimiento es abierto si cualquiera es libre para acceder a él, usarlo, modificarlo y compartirlo bajo condiciones que, como mucho, preserven su autoría y su apertura" (Dutta, 2015; OpenDefinition, 2017).

El Open Access (OA) es un tipo de acceso, no es un tipo de modelo de negocios, de licencia o de contenidos. Las obras disponibles en OA son digitales, gratis, están en línea y mayormente libres restricciones de uso (Suber, 2012). Los repositorios que adoptan este esquema son de libre acceso, lo cual favorece que los motores de búsqueda y rastreadores de información pueden indizar el contenido de estos, permitiendo el acceso no sólo a los metadatos como es el caso de la iniciativa OAI-PMH basada en la norma Dublin Core / ISO 15386 (Ginsparg, Luce, & Van de Sompel, 1999; Van de Sompel & Lagoze, 2000), sino también al contenido completo del material digital. La OAI y el OA son iniciativas separadas que se complementan, sin embargo no deben ser confundidas una con otra.

---

[2] http://roar.eprints.org/

Siempre que se intenta medir la calidad de algo se debe tener en cuenta qué es lo que se define como calidad y cómo se puede obtener una medida de ésta. A veces resulta complicado separar la subjetividad en medidas cualitativas, por ello (Guy, Powell, & Day, 2004) definen la calidad dentro del contexto de los metadatos como: "los metadatos de alta calidad son los que respaldan los requerimientos funcionales del sistema que esté diseñado a soportar", lo que puede resumirse como "la calidad está relacionada con la aptitud para el propósito". Son numerosos y variados los estudios realizados para medir la calidad en los metadatos; por su parte (Park, 2009) realiza un análisis del estado de la investigación y práctica sobre los metadatos basándose en la funcionalidad de éstos, la medición y criterios de evaluación, junto con mecanismos para mejorar la calidad de los mismos, la exactitud, integridad y consistencia son los criterios más comunes utilizados en la medición de metadatos. En (Hillmann, 2008) se presenta un esquema de evaluación de la calidad basada en siete características: integridad, exactitud, procedencia, conformidad con las expectativas, la consistencia lógica y coherencia, oportunidad y accesibilidad. Cada una de ellas cuenta con una serie de criterios e indicadores de cumplimiento que el responsable debe tenerlos en cuenta al momento de realizar la evaluación.

Por otro lado (Ochoa & Duval, 2006) proponen un conjunto de métricas basadas en los mismos parámetros de calidad utilizados por un humano al realizar una revisión de los metadatos. La mayoría de estos estudios analizan la calidad en términos de la generación de los metadatos haciendo revisiones de las técnicas y herramientas implicadas (Guy, Powell, & Day, 2004; Elings & Waibel, 2007; Hillmann, 2008; Man, Wei, Gang, & Juntao, 2010; Mendes, Mühleisen, & Bizer, 2012; Ochoa & Duval, 2009; Clair, 2016), mientras que sólo algunos se centran en la calidad de los datos una vez realizada la recolección de los mismos (Ward, 2002; Shreeves, Knutson, Stvilia, Palmer, Twidale, & Cole, 2005; Nichols, Chan, Bainbridge, Mckay, & Twidale, 2008; Jackson, Han, Groetsch, Mustafoff, & Cole, 2008; Cechinel, Sánchez Alonso, & Sicilia, 2009). Otro tipos de estudios por ejemplo se encargaron de examinar la calidad desde diferentes puntos de vista, analizando los metadatos de repositorios, el cambio y mejora a través del paso del tiempo (Zavalina, Shakeri, & Kizhakkethil, 2015; Palavitsinis, Manouselis, & Sanchez-Alonso, 2017; Marc, 2016), o la interoperabilidad de los metadatos en el tiempo (Sugimoto, Li, Nagamori, & Greenberg, 2016). Sin embargo se echa en falta la existencia de herramientas automáticas o un modelo de datos común que sea interoperable entre los repositorios digitales, para reducir la brecha entre lo que se registra y lo que debe registrarse.

Al igual que el trabajo previo de (Medrano, Figuerola, & Alonso Berrocal, 2012), en esta ocasión se realizó un análisis de los metadatos recolectados para examinar la calidad en términos de consistencia y completitud de los datos sin tener en cuenta el modo en que éstos fueron generados.

Una vez que los datos son recolectados deben ser interpretados y revisados; por ello un análisis de metadatos (Tennant, 2004) debería ser capaz de dar respuesta a preguntas como qué campos de los metadatos están realmente presentes y en qué porcentaje del total de registros, el grado de normalización y consistencia entre los valores reales de esos campos, o si es posible detectar patrones en los contenidos de tales metadatos. Un estudio realizado por (Ward, 2002) sobre 82 repositorios institucionales, indica la baja utilización de los quince elementos del esquema Dublin Core. Un resultado interesante al que llegó el autor fue que existe una media de ocho campos cumplimentados por registro. Más tarde (Shreeves, Knutson, Stvilia, Palmer, Twidale, & Cole, 2005) y las buenas prácticas publicadas por el *Institute of Museum and Library Services Digital Collections and Content Project* (IMLS DCC), confirmarían esos resultados identificando además ocho de estos quince elementos como importantes para la integridad de un registro de metadatos y de utilidad para las búsquedas. Los elementos en cuestión son: *title, creator, subject, description, date, format, identifier* y *rights*.

Materiales y Metodología

El ROAR (Registry of Open Access Repositories) es una de las bases de datos de repositorios de libre acceso más grandes que existen, posee alrededor de 4446 repositorios registrados, con 47 repositorios pertenecientes a instituciones argentinas. El interés fue puesto en los repositorios que soportan el protocolo OAI-PMH, puesto que es un protocolo estándar para la recolección de metadatos y es el objeto de estudio de este trabajo, de existir otros repositorios con otros formatos de recolección no han sido tenidos en cuenta en el presente trabajo. Es necesario aclarar que de estos 47 repositorios la gran parte son repositorios en sí mismos y otros meta-repositorios, es decir, repositorios que agregan contenido de otros repositorios, sin embargo en este estudio no se realiza distinción alguna pues todos son repositorios de libre acceso que poseen metadatos para ser recolectados y susceptibles de ser analizados. De estos 47 repositorios, algunos por el paso del tiempo han dejado de funcionar, ya que ni por la URL del repositorio ni por el proxy provisto por ROAR es posible recuperar los registros, otros simplemente no soportan el protocolo mencionado, con lo cual luego de la verificación de cada uno y la posterior recolección, se obtuvieron 275162 registros correspondientes a 26 repositorios distintos entre el 13-03-2017 y 23-03-2017. La fecha del elemento registrado más antiguo corresponde al año 1794, este campo es utilizado en la mayoría de los casos como la fecha de publicación del elemento digital. El protocolo OAI-PMH permite aplicar ciertos filtros para la recolección de información (Lagoze & Sompel, 2015), sin embargo para este estudio se realizó la recolección completa de todos los repositorios sin aplicar ningún filtro, salvo que el esquema de recolección fuese el *oai_dc*.

El protocolo OAI-PMH propone unos lineamientos generales tanto para listar y recuperar (cosechar) metadatos de un repositorio (OAI Service Providers), como también para exponer recursos (OAI Data Providers) para que puedan ser cosechados por aplicaciones externas. Estos lineamientos proponen la organización de los recursos en conjuntos (sets), el uso del estándar XML para la representación y transporte de recursos (vía HTTP), y un conjunto de seis verbos necesarios para interactuar, como por ejemplo identificar el repositorio, listar conjuntos, listar formatos de metadatos soportados u obtener registros. El mencionado protocolo OAI-PMH es probablemente el más utilizado para brindar interoperabilidad desde el repositorio hacia el exterior (De Giusti, 2016).

Los registros fueron recolectados haciendo uso del protocolo OAI-PMH v 2.0, consultando el esquema de metadatos *oai_dc*[3,4], los repositorios pueden proveer interfaces para otros esquemas, pero el esquema nombrado no solo es el esquema por defecto sino es el obligatorio para cumplir con el protocolo. Para ello se desarrolló una aplicación *harvester* en lenguaje C# bajo el framework de .NET 4.6 utilizando la librería .NET OAI Harvester[5] como medio de recolección de metadatos de los repositorios. Los metadatos fueron almacenados en una base de datos SQL Server 2012 Developer Edition siguiendo el esquema Dublin Core. En total, por cada registro, se almacenaron 18 campos, 3 correspondientes al encabezado (*identifier*, identificador único del registro; *datestamp*, fecha del último acceso al registro y *setSpec*, identificador del conjunto al que pertenece el registro) y los 15 restantes a los metadatos (*title, creator, subject, description, publisher, contributor, date, type, format, identifier, source, language, relation, coverage,* y *rights*).

Resultados y Discusión

Los repositorios argentinos no son de gran tamaño, el 50% de todos los registros se concentra en los dos repositorios más grandes, Consejo Latinoamericano de Ciencias Sociales CLACSO[6] (87970 registros recolectados 31% del total) y SEDICI repositorio institucional de la UNLP[7] (54374 registros 19 % del total). Por otro lado el otro 50% se concentra en 24 repositorios de menor tamaño (menos de 49.000 registros).

En cuanto a la completitud, solo un repositorio del conjunto total, posee los quince campos cumplimentados, el repositorio es la "Escuela Argentina de Tantra"[8], solo que el dato no es tan significativo pues es el repositorio más pequeño con sólo 5 registros. En este mismo sentido el segundo repositorio más completo es Argos[9] Repositorio Institucional de la Secretaría de Investigación y Postgrado de la Facultad de Humanidades y Ciencias Sociales de la Universidad Nacional de Misiones, este cuenta con 511 registros y un 87% de completitud. El tercer repositorio más completo es Biblioteca Digital UNCuyo [10] SID (Sistema integrado de Documentación), Universidad Nacional de Cuyo (UNCuyo), contando con 5371 registros disponibles y un 80% de cumplimiento de la norma.

En cuanto a la cumplimentación de los campos, ya se había adelantado que los 8 campos mayormente cumplimentados eran: *identifier, type, title, date, subject, creator, language* y *description.* En la Figura **1** se ofrece el detalle del porcentaje relativo de cuanto cumplen con el llenado de los campos cada repositorio, esto es, el porcentaje de los registros que cumplen dicho campo dentro de ese repositorio.

---

[3] http://www.openarchives.org/OAI/2.0/OAI-PMH.xsd
[4] http://www.openarchives.org/OAI/2.0/oai_dc.xsd
[5] https://sourceforge.net/projects/netoaihvster/
[6] http://biblioteca.clacso.edu.ar/
[7] http://sedici.unlp.edu.ar/
[8] http://tantra.org.ar/biblioteca/index.php/Biblioteca
[9] http://argos.fhycs.unam.edu.ar/
[10] http://bdigital.uncu.edu.ar/

| Repositorio | title | creator | subject | description | publisher | contributor | date | type | format | source | language | relation | coverage | rights | identifier2 |
|---|---|---|---|---|---|---|---|---|---|---|---|---|---|---|---|
| ESCUELA ARGENTINA DE TANTRA | 100 | 100 | 100 | 100 | 100 | 0 | 100 | 100 | 100 | 100 | 100 | 100 | 100 | 100 | 100 |
| Fadu (Facultad de Arquitectura, Diseño y Urbanismo) | 100 | 100 | 98 | 98 | 0 | 0 | 100 | 100 | 99 | 0 | 0 | 100 | 0 | 0 | 100 |
| RepoCLACAI | 100 | 98 | 98 | 90 | 66 | 1 | 100 | 93 | 0 | 0 | 99 | 23 | 0 | 0 | 100 |
| Repositorio Digital Institucional José María Rosa | 100 | 100 | 100 | 100 | 95 | 1 | 100 | 100 | 100 | 0 | 99 | 0 | 0 | 100 | 100 |
| Balseiro | 99 | 100 | 100 | 98 | 6 | 0 | 97 | 100 | 100 | 0 | 100 | 100 | 0 | 0 | 100 |
| Repositorio Digital San Andrés | 100 | 100 | 86 | 100 | 2 | 1 | 100 | 100 | 0 | 0 | 100 | 0 | 0 | 100 | 100 |
| Repositorio de Ciencias Agropecuarias y Ambientale | 100 | 100 | 12 | 46 | 98 | 45 | 100 | 100 | 100 | 0 | 0 | 100 | 0 | 0 | 100 |
| RPsico (Repositorio de la Facultad de Psicología) | 79 | 77 | 78 | 74 | 1 | 69 | 79 | 79 | 0 | 0 | 79 | 0 | 0 | 13 | 79 |
| Argos | 100 | 100 | 99 | 99 | 100 | 0 | 100 | 100 | 100 | 11 | 99 | 91 | 87 | 100 | 100 |
| Biblioteca de tesis | 100 | 100 | 100 | 100 | 0 | 100 | 100 | 100 | 100 | 0 | 100 | 0 | 0 | 100 | 100 |
| Repositorio Digital de la Universidad FASTA (REDI) | 100 | 99 | 62 | 79 | 1 | 44 | 100 | 100 | 82 | 38 | 85 | 0 | 4 | 98 | 100 |
| Producción Académica UCC | 100 | 100 | 100 | 78 | 32 | 2 | 93 | 100 | 98 | 0 | 98 | 100 | 0 | 83 | 100 |
| Naturalis | 100 | 100 | 100 | 13 | 24 | 14 | 100 | 100 | 100 | 62 | 100 | 12 | 0 | 100 | 100 |
| Filo Digital | 100 | 94 | 74 | 92 | 97 | 68 | 100 | 100 | 99 | 27 | 100 | 27 | 0 | 91 | 100 |
| Nülan | 100 | 98 | 79 | 87 | 44 | 18 | 100 | 99 | 86 | 74 | 86 | 100 | 0 | 86 | 100 |
| FAUBA Digital | 100 | 100 | 89 | 99 | 100 | 30 | 100 | 100 | 100 | 70 | 34 | 0 | 0 | 100 | 100 |
| Universidad Nacional de Córdoba (RD-UNC) | 100 | 98 | 95 | 97 | 45 | 39 | 89 | 99 | 11 | 0 | 100 | 3 | 11 | 100 | 100 |
| Biblioteca Digital Universidad Católica Argentina | 99 | 95 | 99 | 89 | 82 | 38 | 99 | 99 | 99 | 80 | 99 | 0 | 28 | 99 | 99 |
| CIC Digital | 100 | 100 | 100 | 100 | 29 | 20 | 100 | 0 | 100 | 0 | 100 | 0 | 0 | 100 | 100 |
| Biblioteca Digital UNCuyo | 95 | 91 | 99 | 92 | 97 | 28 | 90 | 98 | 100 | 71 | 100 | 7 | 9 | 100 | 100 |
| Biblioteca Digital FCEN-UBA | 100 | 81 | 52 | 77 | 93 | 51 | 99 | 100 | 100 | 33 | 93 | 3 | 7 | 56 | 100 |
| CONICET Digital | 100 | 100 | 100 | 100 | 99 | 0 | 100 | 100 | 100 | 2 | 100 | 99 | 0 | 100 | 100 |
| Repositorio Institucional del Ministerio de Educac | 100 | 61 | 98 | 40 | 28 | 0 | 100 | 99 | 56 | 0 | 64 | 75 | 36 | 0 | 100 |
| Memoria Académica | 100 | 99 | 98 | 63 | 63 | 8 | 100 | 100 | 90 | 92 | 93 | 49 | 0 | 90 | 100 |
| SeDiCI (Servicio de Difusión de la Creación Intele | 100 | 100 | 100 | 95 | 3 | 12 | 100 | 100 | 0 | 2 | 100 | 86 | 0 | 85 | 100 |
| Consejo Latinoamer. de Ciencias Sociales CLACSO | 12 | 11 | 12 | 7 | 12 | 0 | 11 | 95 | 12 | 0 | 11 | 8 | 10 | 12 | 100 |

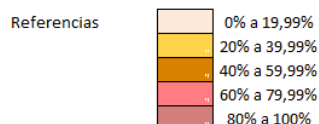

Referencias:
- 0% a 19,99%
- 20% a 39,99%
- 40% a 59,99%
- 60% a 79,99%
- 80% a 100%

Figura 1: Porcentaje relativo de cumplimentación de cada repositorio

Por otro lado la **Tabla 1** ofrece un resumen del porcentaje de cumplimentación absoluto de los 15 campos Dublin Core, esto es, el porcentaje total de todos los registros de todos los repositorios que cumplen cierto campo, se agregó a los quince elementos el campo *setSpec*, si bien este campo pertenece al encabezado y no a los metadatos, resulta también de interés tenerlo en cuenta, pues en este caso está cumplimentado al 100%.

Tabla 1: Porcentaje absoluto de cumplimentación de los campos

| Atributo | Porcentaje |
|---|---|
| setSpec | 100,00 |
| identifier2 | 99,95 |
| type | 96,32 |
| title | 71,69 |
| date | 71,21 |
| subject | 69,00 |
| creator | 65,91 |
| language | 64,75 |
| description | 53,40 |
| rights | 52,93 |
| relation | 43,45 |
| format | 42,97 |
| publisher | 32,21 |
| source | 22,20 |
| contributor | 8,82 |
| coverage | 8,61 |

El campo *identifier* de los metadatos casi es completado al 100%, el pequeño faltante se debe a que los repositorios RPsico[11] (Repositorio de la Facultad de Psicología, Universidad Nacional de Mar del Plata) y Biblioteca Digital Universidad Católica Argentina[12] lo cumplimentan en un 79% y 99% respectivamente, con 358 y 4741 registros cada uno. El campo *title* también es interesante, a pesar del alto porcentaje presentado, que un registro o publicación almacenado no posea el título si es significativo y más aún si en el repositorio de mayor tamaño (Consejo Latinoamericano de Ciencias Sociales CLACSO, 87970 registros) solo lo cumplimentan el 12% de los registros. Los otros repositorios que no lo cumplimentan al 100% lo hacen al 79% (RPsico), 95% (Biblioteca Digital UNCuyo) y 99% (Repositorio Institucional del Centro Atómico Bariloche y el Instituto Balseiro y Biblioteca Digital Universidad Católica Argentina). El resto de repositorios (21 en total), cumplen con este campo al 100%. Los campos menos cumplimentados (menos del 50%) no son críticos salvo el campo *format*, este campo puede resultar útil para las tareas de recuperación en los casos que sea necesario acceder al objeto catalogado, ya que dependiendo del formato se podrá automatizar el almacenamiento o procesamiento del recurso digital.

Del total de registros que cumplimentan el campo *description* (146946 registros), el 0,1% de los registros posee longitud superior a 10000 caracteres, el 50% entre 1000 y 10000 caracteres y el 49% menos de 1000 caracteres. Las tres longitudes con mayor frecuencia son: 122 caracteres con 910 registros, 125 caracteres con 712 registros y 119 caracteres con 591 caracteres.

Para los registros que cumplimentan el campo title (casi el 72% del total), el 68% posee longitud entre 1-100 caracteres, el 26% entre 101-200 caracteres, el 4% entre 201-300 caracteres y el 1% entre 301-400 caracteres. Las tres longitudes con mayor frecuencia son: 12 caracteres con 4272 registros, 71 caracteres con 1882 caracteres y 18 caracteres con 1861 registros.

Si bien la calidad es subjetiva y relativa a lo que se está midiendo, en este caso se puede evaluar la calidad en términos de la normalización de los campos más representativos: códigos o abreviaturas de lenguas (campo *language*), normalización y utilidad de palabras claves (campo *subject*), tipo de publicación (campo *type*), el formato de las mismas (campo *format*) y normalización de nombres personales (campos *creator*).

El campo *language* indica el lenguaje en el que está escrita la publicación, en los datos recolectados existen 91 formas distintas para referirse a este campo, sin embargo, como muestra la Tabla **2**, las 10 variantes más representativas cubren el 64,41% de los registros, mientras que el 35,25% son cadenas vacías y el 0,09% corresponde al resto de las variantes.

---

[11] http://rpsico.mdp.edu.ar/
[12] http://bibliotecadigital.uca.edu.ar/greenstone/cgi-bin/library.cgi

Tabla 2: Variantes más representativas del campo Language

| Variante del campo Language | Cantidad de registros | Porcentaje sobre el total de registros |
|---|---:|---:|
| spa | 78314 | 28,46 |
| es | 71674 | 26,05 |
| eng | 8450 | 3,07 |
| Español | 6837 | 2,48 |
| en | 4055 | 1,47 |
| spa;spa | 2808 | 1,02 |
| por | 1842 | 0,67 |
| pt | 1380 | 0,50 |
| es;spa | 1337 | 0,49 |
| Inglés | 555 | 0,20 |
| fr | 86 | 0,03 |
| fre | 80 | 0,03 |
| fra | 73 | 0,03 |
| Por;Spa | 72 | 0,03 |
| Spa;Por | 72 | 0,03 |
| eng;eng | 68 | 0,02 |
| Portugués | 60 | 0,02 |
| spa;eng | 57 | 0,02 |
| ita | 48 | 0,02 |
| spa;spa;spa;spa | 29 | 0,01 |
| eng;spa | 27 | 0,01 |
| otros | 252 | 0,09 |
| **vacíos** | **96986** | **35,25** |

La utilización de descriptores, palabras claves o *keywords*, son especificadas en el campo *subject*. El 69% de los registros (189870) posee al menos un descriptor. Este campo es muy particular, pues es sabido que un registro se encuentra bien "descripto" cuando la utilización de estos elementos se hace de forma medida y cuidadosa, es decir, no es una decisión al azar la elección de estos elementos para describir un registro, pues gran parte de las búsquedas y procesos de recolección de información se realizan a partir de estas palabras. Existen alrededor de 814999 descriptores distintos en todo el conjunto de registros, por lo que salta a la vista que un registro puede utilizar más de uno, es más, el registro con la mayor cantidad de descriptores posee 38 elementos, el promedio general es de 5 descriptores por registro. El Top 10 de los descriptores más utilizados lo encabezan: Ciencias Informáticas (utilizado 9356 veces), Humanidades (7339 veces), EDUCACION SUPERIOR (6602), Educación (6157), HISTORIA (5969), RECONOCIMIENTO OFICIAL DE TITULOS (5931), PLAN DE ESTUDIOS (5806), enseñanza universitaria (5571), CIENCIAS NATURALES Y EXACTAS (5458) y Literatura (4452). En la Figura **2** se pueden observar el nivel de utilización de descriptores tanto en el título como en la descripción de los registros almacenados. Para entender mejor este gráfico, en el eje x se encuentran las distintas cantidades de descriptores, como ejemplo, de los registros que poseen 6 descriptores (18715 registros), existen 11911 registros que incluyen al menos uno de estos descriptores en el título y existen 13398 registros que incluyen al menos uno de los descriptores en la descripción.

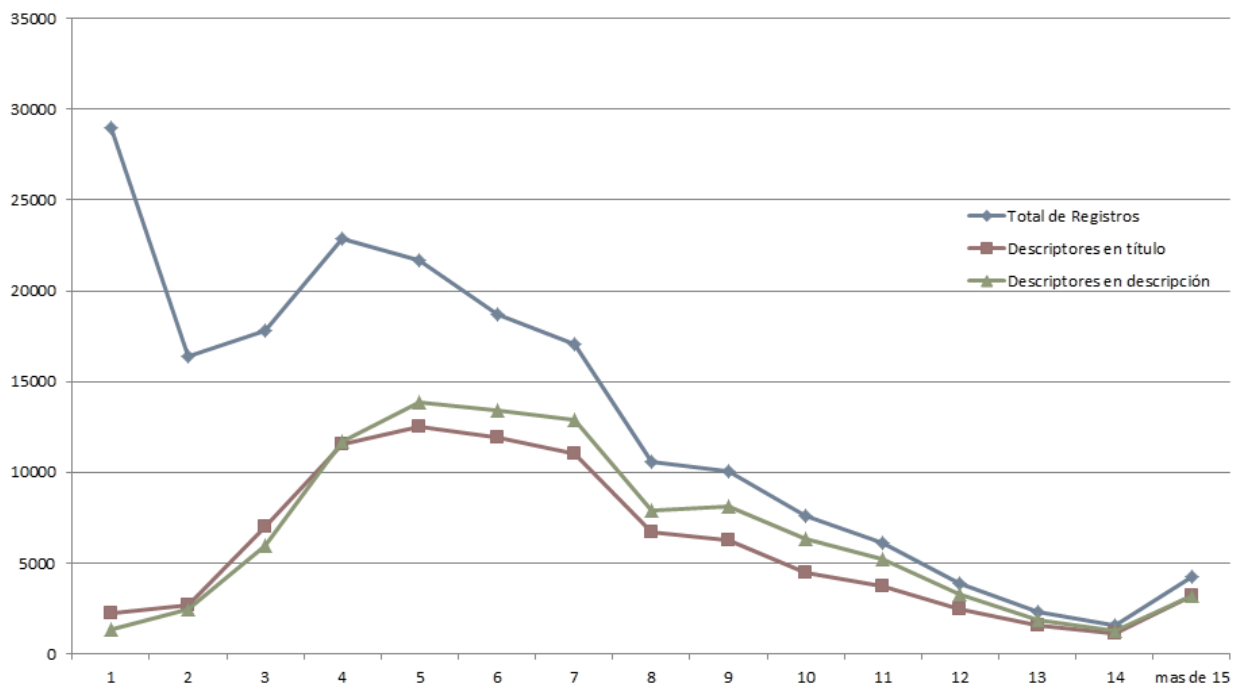

Figura 2: Cantidad de Descriptores en Títulos y Descripciones

El campo *type* señala de qué tipo de publicación se trata, para este trabajo se encontraron 1056 formas distintas de especificar este campo, además se encuentra cumplimentado en un 96% sobre el total de los registros (el 4% son cadenas vacías), por otro lado esta gran diversidad demuestra la enorme falta de normalización en este campo, es más, existen alrededor de 443 formas distintas para referirse a un "Artículo", aún más alarmante es el hecho que de estas 443 formas, 364 se encuentren en el repositorio CLACSO. A nivel global, entre las formas más comunes se encuentran: "*article*", "*conferenceObject*", "*Objeto de conferencia*", "*Artículo científico*", "*Articulo;Articulo*", "*Article*" y "*legislation*". Como se observa en la **Tabla 3** (se listan las 20 versiones más frecuentes de este campo junto al total de campos vacíos y agrupados como "otros" el resto de las variantes), estas versiones listadas cubren el 72% del total de registros recolectados.

Tabla 3: Variantes más representativas del campo Type

| Variantes del campo type | Cantidad | Porcentaje |
| --- | --- | --- |

| | de registros | sobre el total de registros |
|---|---|---|
| info:eu-repo/semantics/article;info:ar-repo/semantics/artículo;info:eu-repo/semantics/publishedVersion | 27320 | 9,93 |
| info:eu-repo/semantics/conferenceObject;info:ar-repo/semantics/documento de conferencia;info:eu-repo/semantics/publishedVersion | 23948 | 8,70 |
| info:eu-repo/semantics/article;info:eu-repo/semantics/publishedVersion | 23245 | 8,45 |
| Objeto de conferencia;Objeto de conferencia | 19775 | 7,19 |
| Artículo científico | 19061 | 6,93 |
| Articulo;Articulo | 18089 | 6,57 |
| Article | 12818 | 4,66 |
| legislation | 12740 | 4,63 |
| text | 7258 | 2,64 |
| Imagen | 4619 | 1,68 |
| info:eu-repo/semantics/review;info:ar-repo/semantics/revisión literaria; info:eu-repo/semantics/publishedVersion | 4108 | 1,49 |
| article;info:ar-repo/semantics/artículo;info:eu-repo/semantics/article;info:eu-repo/semantics/publishedVersion | 3724 | 1,35 |
| Tesis;Tesis de doctorado | 3589 | 1,30 |
| info:eu-repo/semantics/article;info:eu-repo/semantics/publishedVersion;Artículo revisado por pares | 2993 | 1,09 |
| Reseña | 2981 | 1,08 |
| Text;draft;Capítulo de Libro | 2951 | 1,07 |
| info:eu-repo/semantics/bachelorThesis;info:ar-repo/semantics/tesis de grado;info:eu-repo/semantics/acceptedVersion | 2874 | 1,04 |
| info:eu-repo/semantics/article;artículo;info:eu-repo/semantics/publishedVersion | 2424 | 0,88 |
| Artículo | 2403 | 0,87 |
| Articulo;Revision | 2136 | 0,78 |
| otros | 65987 | 23,98 |
| **vacíos** | **10119** | **3,68** |

El campo utilizado para indicar el formato del elemento catalogado es *format*, cumplimentado casi un 43%, posee la alarmante cifra de 14142 formas distintas. Siendo la predominante la cadena "application/pdf" con casi el 25% del total de los registros, sin embargo existen 4281 formas distintas para referirse al formato "PDF" donde las más frecuentes son: "application/pdf", "text/html;application/pdf", "text; pdf", "pdf", "application/pdf;application/pdf", "application/pdf;4 p.", "application/pdf;6 p.", entre tantas otras miles, alcanzando entre todas estas el 35% del total de los registros. En cambio para el formato "HTML" solo existen tres formas distintas ("application/html", "text/html",

"text/html;application/pdf", donde la última hace referencia también al formato PDF) representando estas solo el 4% de los registros almacenados.

En cuanto al campo *creator*, que es el utilizado para señalar los autores o creadores del elemento digital, el cual es cumplimentado por casi el 66% de los registros, contiene 135728 autores distintos, de estos, el 92% se encuentra bajo la forma "*Apellidos, Nombres*", mientras que el porcentaje restante en la forma "*Nombres + Apellidos*". La **Tabla 4** resume la cantidad de autores por publicación, con la etiqueta "+ de 10" se indican los registros que poseen entre 11 y 32 autores que es el número máximo encontrado.

Tabla 4: Cantidad de Autores por registros

| Cantidad de autores | Cantidad de registros |
|---|---|
| 1 | 122020 |
| 2 | 22711 |
| 3 | 14046 |
| 4 | 8611 |
| 5 | 4855 |
| 6 | 3181 |
| 7 | 1825 |
| 8 | 1199 |
| 9 | 764 |
| 10 | 553 |
| + de 10 | 1587 |

Conclusiones

Según datos del año 2015 de la Secretaría de Políticas Universitarias del Ministerio de Educación y Deportes de la Nación (Universitarias, 2017), el sistema universitario Argentino está formado por 53 Universidades Nacionales, 49 Universidades Privadas, 7 Institutos Universitarios Estatales, 14 Institutos Universitarios Privados, 6 Universidades Provinciales, 1 Universidad Extranjera y 1 Universidad Internacional, de estas instituciones sólo existen 47 repositorios registrados en ROAR y de estos solo 26 repositorios estén disponibles y accesibles para la recolección de información. Sobre este conjunto de repositorios, en este trabajo, se lograron identificar problemas comunes a la mayoría de ellos y relacionados con la normalización de datos, principalmente títulos y descriptores muy extensos, un enorme número de autores, diferentes formas de referirse al mismo lenguaje o al mismo tipo de archivo. Hablando específicamente de los metadatos, se podrían implementar mecanismos para reemplazar y controlar la formación del campo *language*, los registros se pueden encuadrar en no más de 10 idiomas, lo mismo sucede con el campo *type* y con la innecesaria inclusión de palabras extras para referirse a un tipo de publicación. Quizás simplemente delimitando el largo de este campo se pueda reducir este desfasaje.

El campo *format* también presenta un mal uso pues la inclusión de formatos ficticios o la redundancia en repetir el nombre o combinaciones de estos es verdaderamente alarmante. A medida que aumenta la cantidad de descriptores incluidos en los metadatos también aumenta la cantidad de estos elementos utilizados en títulos y descripciones, este comportamiento casi es lineal y esto no es un buen indicador de calidad, ya que la función del descriptor es centrar el contenido del elemento catalogado en unas pocas palabras y no en frases u oraciones completas para intentar dar "información de calidad", para este campo sería ideal contar con listados de palabras de acceso público y de uso frecuente, o revisar las tareas o herramientas encargadas de la catalogación. Que un recurso no posea los datos de su creador o autor es más que significativo, que el 34% de los registros no posea este dato habla a las claras del poco control que existe en este tipo de repositorios.

Es de destacar que si bien la especificación del formato Dublin Core no exige que los campos se cumplimenten en un formato determinado, ya que son simples cadenas de caracteres, si propone normas y esquemas que los proveedores de datos pueden implementar para facilitar la interoperabilidad, ejemplo de ello es la aplicación de las directrices DRIVER 2.0[13] u otras basadas en esta (OpenAIRE[14] y SNRD[15]) cuyo objetivo es la normalización de la representación de algunos metadatos y el cumplimiento de ciertos metadatos de forma obligatoria, recomendada u opcional, esta estandarización en la representación y codificación de metadatos, además de los protocolos de comunicación, permite la interoperabilidad entre los distintos sistemas de información. Contar con esquemas formales y revisiones periódicas del contenido almacenado en estos repositorios, aumentaría notablemente la calidad de la información y del servicio que brindan, sobre todo las tareas de recolección se verían enormemente favorecidas con estos pequeños cambios.

---

[13] http://travesia.mcu.es/portalnb/jspui/handle/10421/1441
[14] https://guidelines.openaire.eu/en/latest/
[15] http://repositorios.mincyt.gob.ar/pdfs/Directrices_SNRD_2013.pdf